\documentclass{PoS}
\usepackage{epsfig}
\usepackage{amsmath}
\usepackage{float}
\usepackage{graphicx}
\usepackage[figuresright]{rotating}

\def\td#1{\widetilde{\delta}\left(#1\right)}
\def\thd#1{\tilde{\theta}\left(#1\right)}

\def\nn{\nonumber}

\title{  {\footnotesize IFIC/10-44} \\
          Feynman's Tree Theorem and Loop-Tree Dualities}

\ShortTitle{Tree-Loop Duality}

\author{Isabella Bierenbaum \footnote{As of October 2010: Institut f\"ur Theoretische Teilchenphysik und Kosmologie
        RWTH Aachen D-52056 Aachen, Germany. \; E-mail: bierenbaum@physik.rwth-aachen.de}\\
        Instituto de F\'{\i}sica Corpuscular, Universitat de Valencia-Consejo Superior de Investigaciones Cient\'{\i}ficas,
	Apartado de Correos 22085, E-46071, Valencia, Spain\\
        E-mail: \email{isabella.bierenbaum@ific.uv.es}}

\author{Stefano Catani\\
        INFN, Sezione di Firenze and Dipartimento di Fisica, Universita di Firenze, I-50019 Sesto Fiorentino, Florence, Italy\\
        E-mail: \email{stefano.catani@fi.infn.it}}
		
\author{Petros Draggiotis\\
        Instituto de F\'{\i}sica Corpuscular, Universitat de Valencia-Consejo Superior de Investigaciones Cient\'{\i}ficas,
	Apartado de Correos 22085, E-46071, Valencia, Spain\\
        E-mail: \email{Petros.Drangiotis@ific.uv.es}}

\author{\speaker{German Rodrigo}\\
        Instituto de F\'{\i}sica Corpuscular, Universitat de Valencia-Consejo Superior de Investigaciones Cient\'{\i}ficas,
	Apartado de Correos 22085, E-46071, Valencia, Spain \\
        E-mail: \email{german.rodrigo@ific.uv.es}}

\abstract{We discuss the duality theorem, which provides a relation between loop integrals and phase space integrals. 
          We rederive the duality relation for the one-loop case and extend it to two and higher-order loops.
	  We explicitly show its application to two- and three-loop scalar master integrals and discuss the structure
	  of the occurring cuts.}

\FullConference{Light Cone 2010 - LC2010\\
		June 14-18, 2010\\
		Valencia, Spain}

\begin{document}

\section{Introduction}
In recent years, important efforts have been devoted to developing efficient methods for
the calculation of multi--leg and multi--loop diagrams, realized, e.g., by
unitarity--based methods or by traditional Feynman diagram approaches,
\cite{Binoth:2010ra}.

The computation of cross sections at next-to-leading order (NLO) 
or next-to-next-to-leading order (NNLO) requires the separate
evaluation of real and virtual radiative corrections, which are given in the
former case by multi--leg tree--level and in the latter by multi--leg loop
matrix elements to be integrated over the multi--particle phase--space of the
physical process.  The loop--tree duality relation at one--loop presented in
Ref.~\cite{Catani:2008xa}, as well as other methods relating one--loop and
phase--space integrals, \cite{Soperetal,Kilian:2009wy,Moretti:2008jj}, recast
the virtual radiative corrections in a form that closely parallels the
contribution of the real radiative corrections. The use of this close
correspondence is meant to simplify calculations through a direct combination
of real and virtual contributions to NLO cross sections.  Furthermore, the
duality relation has analogies with the Feynman Tree Theorem (FTT),
\cite{Feynman:1963ax,F2}, but offers the advantage of involving only single
cuts of the one--loop Feynman diagrams.

In this talk we will present the equivalence of the FFT and duality theorems at
one-loop level and extend it to higher order loops, with the aim of applying it to
NLO and NNLO cross sections.

\section{The Duality Theorem at One Loop}
We start by considering a general one--loop $N$--leg diagram, as
shown in Fig. 1., which is represented by the scalar integral:
\begin{equation}
L^{(1)}(p_1, p_2, \dots, p_N) =
\int_{\ell_1} \, \prod_{i=1}^{N} \, G_F(q_i)~.
\end{equation}
The four--momenta of the external legs are denoted $p_i$, $i \in
\{1,2,\ldots N\}$. All are taken as outgoing and ordered clockwise.
We use dimensionally regularized integrals with the number of space--time 
dimensions equal to $d$, and introduce the following shorthand notation:
\begin{equation}
\int_{\ell_i} \, \cdots  \equiv
- i \, \int \frac{d^d \ell_i}{(2\pi)^d} \, \cdots~.
\end{equation}
The FFT and the duality theorem, both depend on the pole structure of the Feynman and the
advanced propagators. These are defined as
\begin{equation}
G_F(q_i) = \frac{1}{q_i^2-m_i^2+i0}\;, \;\;\; G_A(q_i) = \frac{1}{q_i^2-m_i^2-i0\,q_{i,0}}~,
\end{equation}
with $q_i$ being the $d$--dimensional four momentum, whose energy (time
component) is $q_{i,0}$. The poles of both kinds of propagators in the complex $q_{i,0}$--plane are placed at:
\begin{equation}
\label{fpole}
q^F_{i,0} = \pm  {\sqrt {{\bf q}_i^2 -m_i^2-i0}} \;\; , \;\; q^A_{i,0} \simeq \pm  {\sqrt {{\bf q}_i^2 -m_i^2}} +i0~,
\end{equation}
respectively. Thus, the pole with positive (negative) energy of the 
Feynman propagator is slightly displaced below (above) the real axis, while both poles
(independent of the sign of energy) of the advanced propagator are slightly displaced 
above the real axis. Using the elementary identity:
\begin{equation}
\frac{1}{x \pm i 0}= PV\left( \frac{1}{x} \right)\mp i \pi \delta (x)~, 
\end{equation}
where $PV$ denotes the principal value, we find the following relation between the
two propagators:
\begin{equation}
\label{AdvFeyn}
G_{A}(q_i) = G_{F}(q_i)+\td{q_i} , \qquad \td{q_i}= 2 \pi \, i \, \theta(q_i^0) \delta (q_i^2-m_i^2) .
\end{equation}
We define the advanced loop integral from the usual 
definition of the loop integral by replacing Feynman with advanced propagators:
\begin{equation}
L_A^{(1)}(p_1,p_2, \ldots , p_N )= \int_{\ell_1} \prod_{i=1}^N G_A (q_i)
\end{equation}
Then by closing a contour on the lower $\ell_{1,0}$--complex-plane and computing the integral with the residue method, we notice that:
\begin{equation}
L_A^{(1)}(p_1,p_2, \ldots , p_N )= 0~,
\end{equation}
since all the poles of the advanced propagators are in the positive half-plane.
Replacing the advanced propagators with Feynman propagators, Eq.(\ref{AdvFeyn}), we arrive at the relation:
\begin{equation}
L^{(1)}(p_1,p_2, \ldots , p_N )=- \left[ L_{1-cut}^{(1)}(p_1,p_2, \ldots , p_N )+ \cdots 
                                       + L_{N-cut}^{(1)}(p_1,p_2, \ldots , p_N )   \right]~. 
\end{equation}
This equation is the FFT at one loop. It relates the one loop integral
to multiple cut integrals. Each delta function $\td{q_i}$ in the multiple cut terms, replaces the 
corresponding Feynman propagator, by cutting the internal line with momentum $q_i$. 

Rather than starting from $L^{(1)}_A$, we apply directly the residue theorem to the computation of $L^{(1)}$. 
Closing the integration contour at infinity in the direction of the negative
imaginary axis, according to the Cauchy theorem, one picks up one pole from
each of the $N$ Feynman propagators. 
\begin{equation}
L^{(1)}(p_1,p_2, \ldots , p_N )=- 2 \pi i \int_{\ell_1} \sum {\rm Res}_{ \left\lbrace {\rm Im} q_{i,0} <0 \right\rbrace } 
\left[ \prod_{j=1}^N G_F (q_j) \right] ~.
\end{equation}
In Ref.~\cite{Catani:2008xa}, it is shown that the residue of these poles is given by:
\begin{equation}
{\rm Res}_{ \left\lbrace {\rm Im} q_{i,0} <0 \right\rbrace } \frac{1}{q^2_i -m_i^2 +i0}= \int d \ell_{1,0} \delta_+ (q_i^2 -m_i^2)~.
\end{equation}
The value of the rest of the propagators at the residue, is shown to be~\cite{Catani:2008xa} :
\begin{equation}
\left[  \prod_{j\neq i} G_F (q_j) \right]_{ \left\lbrace q_i^2-m_i^2 = -i0 \right\rbrace } =
\prod_{j\neq i} \frac{1}{q_j^2 -m_j^2 -i 0 \eta (q_j - q_i )}~,
\end{equation}
where $\eta$ is a future-like vector:
\begin{equation}
\label{etadef}
\eta_\mu = (\eta_0, {\bf \eta}) \;\;, \;\; \quad \eta_0 \geq 0, 
\; \eta^2 = \eta_\mu \eta^\mu \geq 0 ~.
\end{equation}
We finally obtain:
\begin{equation}
\label{DualTheorem}
L^{(1)}(p_1,p_2, \ldots , p_N )= - \sum_{i=1}^N \int_{\ell_1} \td{q_i} \prod_{j \neq i} G_D (q_i ; q_j)~,
\end{equation}
where
\begin{equation}
G_D(q_i;q_j) = \frac{1}{q_j^2 -m_j^2 -i 0 \eta (q_j - q_i )}
\end{equation}
is the \textit{dual} propagator, which depends on the vector $\eta$. This is the duality theorem. 
We notice that since all the internal momenta depend on the same loop
momentum, all dependence on it drops out and the $i0$ prescription depends solely on the external momenta. This is
an important trait and we will try to generalize it, in the next section, to more loops. The presence of the vector
$\eta_\mu$ is a consequence of using the residue theorem and the fact that the
residues at each of the poles are not Lorentz--invariant quantities.  The
Lorentz--invariance of the loop integral is recovered after summing over all
the residues.

Finally we note that using the following relation between dual and Feynman propagators
\begin{equation}
\td{q_i} G_D(q_i;q_j) = \td{q_i} \Bigl[ G_F(q_j) + \thd{q_j-q_i} \; \td{q_j} \Bigr]~,
\end{equation}
where $\thd{q}=\theta (\eta q)$, we can show that FFT and the duality theorem are equivalent \cite{Catani:2008xa}.

\section{Duality theorem at two loops}

\begin{figure*}[t]
\begin{center}
  \epsfig{file=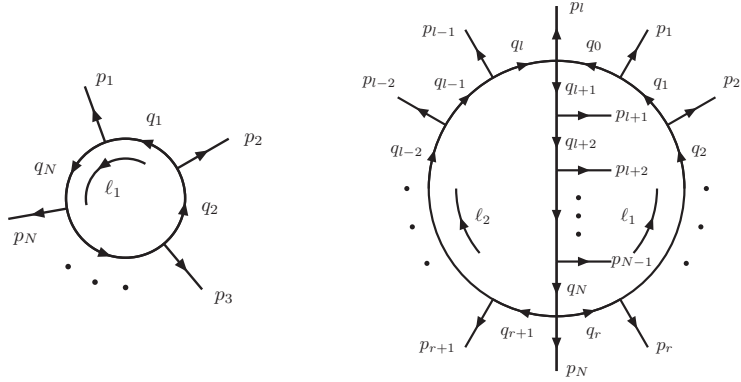,height=5cm}
\end{center}
\vspace*{-6mm}
\caption{\label{f2loop} {\em Momentum configuration of the one--loop and
    two--loop $N$--point scalar integral. Note, that in the two--loop case
    $p_l$ and $p_N$ can be equal to zero and hence the number of internal
    lines can differ from the number of external momenta.}}
\end{figure*}
Our main goal is to write down the extension of the duality theorem in two- and higher
loops keeping the dependence of the $i0$ prescription at the dual propagator, on external momenta only. Also, we want to
formulate it in such a way that we have the same number of cuts as the number of loops.
At two loops the generic graph with $N$-legs is shown in Fig. 1. All momenta are taken outgoing and we
now have two integration momenta $\ell_1 , \, \ell_2$.
To extend the duality theorem beyond one loop, it is useful to extend the definition of propagators
of single momenta, to combinations of propagators of sets of internal momenta. To this end, let $\alpha_k$ be
any set of loop momenta. We define the following functions of Feynman, advanced and dual
propagators:
\begin{eqnarray}
\label{eq:multi}
G_{F(A)}(\alpha_k) &=& \prod_{i \in \alpha_k} G_{F(A)}( q_i)~, \nn \\
G_D(\pm \alpha_k)\; &=& \sum_{i \in \alpha_k} \, \td{\pm q_{i}} \, 
\prod_{\substack{j \in \alpha_k \\ j \neq i }} \, G_D(\pm q_i;\pm q_j)~. 
\end{eqnarray}
The minus sign in the definitions above, signifies a change in the flow of the momenta of
the set. For a single momentum, $\alpha_k = \{i\}$ we define $G_D(\pm \alpha_k)=\td{\pm q_i}$.
The relation between the three kind of propagators (see Eq.(\ref{AdvFeyn})), in the any-loop extended form, is now:
\begin{equation}
G_A(\alpha_k) = G_F(\alpha_k) + G_D(\alpha_k)~.
\label{eq:relevant}
\end{equation}
With the notation described above, the one loop theorem, Eq.(\ref{DualTheorem}) can now be written in the compact form:
\begin{equation}
\label{Ln}
L^{(1)}(p_1, p_2, \dots, p_N) =
\int_{\ell_1}  G_F(\alpha_1) = -  \int_{\ell_1}  G_D(\alpha_1)~.
\end{equation}
When we have the union of several subsets of momenta, $\beta_N \equiv \alpha_1 \cup ... \cup
\alpha_N$ we can write the dual propagator function in terms of these
subsets, by using the multiplicative properties of (\ref{eq:multi}):
\begin{equation}
\label{GAinGDGeneralN}
G_D(\alpha_1 \cup \alpha_2 \cup ... \cup \alpha_N) = 
\sum_{\substack{\beta_N^{(1)} \cup     \beta_N^{(2)}\\  = \beta_N}} \,
\prod_{\substack{i_1\in \beta_N^{(1)}}} \, G_D(\alpha_{i_1}) \,
\prod_{\substack{i_2\in \beta_N^{(2)}}} \,G_F(\alpha_{i_2})\,
\end{equation}
An example, that will be used shortly for the two-loop case, is when $N=2$, e.g. $\beta_2 = \alpha_1 \cup \alpha_2$ :
\begin{equation}
\label{UnionExample}
G_D \left( \alpha_1 \cup \alpha_2 \right) = G_D (\alpha_1)G_D (\alpha_2)+G_D (\alpha_1)G_F (\alpha_2)+
G_F (\alpha_1)G_D (\alpha_2)~.
\end{equation}

Let us now turn to the two loop case. Unlike the one-loop, the number of
external momenta may be different from the number of internal momenta. The momenta of the internal lines 
are again denoted by $q_i$ and are explicitely given by
\begin{eqnarray}
\label{defqi2l}
q_i = \left\{
\begin{tabular}{ll}
$\ell_1+p_{1,i}$ & , $i \in \alpha_1$ \\
$\ell_2+p_{i,l-1}$ & , $i \in \alpha_2$ \\
$\ell_1+\ell_2+p_{i,l-1}$ &  , $i\in \alpha_3$ ~,
\end{tabular}
\right. 
\end{eqnarray}
where $\alpha_k$, with $k=1,2,3$, are defined as the set of lines, propagators
respectively, related to the momenta $q_i$, for the following ranges of $i$:
\begin{equation}
\label{lines}
\alpha_1= \{0,1,...,r\} , \; \; \alpha_2= \{r+1,r+2,...,l\} , \;\; \alpha_3= \{l+1,l+2,...,N\}. 
\end{equation}
The expression for the two-loop $N$-leg scalar integral is:
\begin{equation}
L^{(2)}(p_1, p_2, \dots, p_N)=\int_{\ell_1} \int_{\ell_2} G_F (\alpha_1 \cup \alpha_2 \cup \alpha_3)~.
\end{equation}
Using the duality theorem sequentially, first for the loop momenta $\ell_1$ and the expression (\ref{UnionExample}), we
arrive at the dual representation of the two-loop scalar integral,
\begin{equation}
\label{L2firstloop}
L^{(2)}(p_1, p_2, \dots, p_N) = - \int_{\ell_1} \, \int_{\ell_2} \,
\left\{ G_D(\alpha_1) \, G_D(\alpha_3)
+ G_D(\alpha_1) \, G_F(\alpha_3) + G_F(\alpha_1) \, G_D(\alpha_3) \right\} \, G_F(\alpha_2). 
\end{equation}
The first term of the integrand on the right--hand side of Eq.(\ref{L2firstloop})
is the product of two dual functions, and therefore already contains double
cuts.  We do not modify this term further. The second and third terms of
(\ref{L2firstloop}) contain only single cuts and we thus apply the duality
theorem again, i.e., use (\ref{Ln}) for $\ell_2$. A subtlety arises at
this point since due to our choice of momentum flow, $\alpha_1$ and $\alpha_2$
appearing in the third term of (\ref{L2firstloop}), flow in the opposite
sense. Hence, in order to apply the duality theorem to the second loop, we have
to reverse the momentum flow of one of these two loop lines. We choose to
change the direction of $\alpha_1$, namely $q_i\to -q_i$ for $i \in \alpha_1$. 
Thus, applying (\ref{Ln}) to the last two terms of
(\ref{L2firstloop}) and expanding all parts in terms of the single loop lines
of (\ref{lines}) leads to
\begin{eqnarray}
\label{AdvDualstar}
&&L^{(2)}(p_1, p_2, \dots, p_N)  \\
&=& \int_{\ell_1} \int_{\ell_2} \, \left\{
  G_D(\alpha_1)  \, G_D(\alpha_2) \, G_F(\alpha_3) 
+ G_D(-\alpha_1) \, G_F(\alpha_2) \, G_D(\alpha_3)
+ G^*(\alpha_1)  \, G_D(\alpha_2) \, G_D(\alpha_3) \right\} \nn ~,
\end{eqnarray}
where
\begin{equation}
\label{Gstar1}
G^*(\alpha_k) \equiv G_F(\alpha_k) + G_D(\alpha_k) + G_D(-\alpha_k)~.
\end{equation}
In (\ref{AdvDualstar}), the
$i0$ prescription of all the dual propagators depends on external momenta
only. Through (\ref{Gstar1}), however, (\ref{AdvDualstar}) contains also triple
cuts, given by the contributions with three $G_D(\alpha_k)$. The triple cuts
are such that they split the two--loop diagram into two disconnected
tree--level diagrams. By definition, however, the triple cuts
are such that there is no more than one cut per loop line $\alpha_k$.

\section{Duality theorem at three loops and beyond}

\begin{figure*}[t]
\begin{center}
\epsfig{file=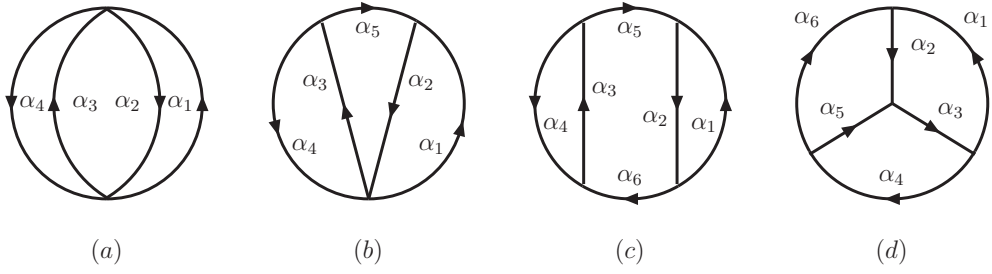,height=3.5cm}
\end{center}
\vspace*{-7mm}
\caption{\label{f3loop} {\em Master topologies of three--loop scalar
    integrals. Each internal line $\alpha_i$ can be dressed with an arbitrary
    number of external lines, which are not shown here.  }}
\end{figure*}

Beyond two loops, the duality theorem applies in a similar manner. We expect to
find at least the same number of cuts as the number of loops, and topology
dependent disconnected tree diagrams built by cutting up to all the loop lines
$\alpha_k$. For the case of three loops
there are four master topologies, shown in Fig. 2. As an example, the basketball graph, Fig. 2a, is in terms of dual 
propagators:
\begin{equation}
\label{basket1}
L^{(3)}_{\rm basket}(p_1, p_2, \dots, p_N) =  
\int_{\ell_1} \int_{\ell_2} \int_{\ell_3} \,
G_D(\alpha_1 \cup \alpha_2)\; G_D(\alpha_3 \cup \alpha_4)~.
\end{equation}
If we expand all existing dual functions in (\ref{basket1}) in terms of 
dual functions of single loop lines by using (\ref{GAinGDGeneralN}) and apply the duality theorem 
to the third loop, we obtain
\begin{eqnarray}
\label{BasketExp}
&& L^{(3)}_{\rm basket}(p_1, p_2, \dots, p_N) 
= -\int_{\ell_1} \int_{\ell_2} \int_{\ell_3} \,
\bigg\{
G_D(\alpha_2,\alpha_3,-\alpha_4)\; G_F(\alpha_1)
+ G_D(\alpha_1,\alpha_3,-\alpha_4)\; G_F(\alpha_2) \nn \\ 
&&\qquad \bigg.
+ G_D(-\alpha_1,\alpha_2,\alpha_4)\; G_F(\alpha_3)
+ G_D(-\alpha_1,\alpha_2,\alpha_3)\; G_F(\alpha_4) \nn \\
&&\qquad \bigg.
+ G_D(-\alpha_1,\alpha_2,\alpha_3,\alpha_4)
+ G_D(\alpha_1,\alpha_2,\alpha_3,-\alpha_4)
+ G_D(-\alpha_1,\alpha_2,\alpha_3,-\alpha_4)
\bigg\}~,
\end{eqnarray}
where for brevity we use the notation: $G_D(\alpha_1, \ldots , \alpha_N)=\prod_{i=1}^N G_D(\alpha_i)$. For more details and the expressions for the rest of the topologies, we refer the reader to \cite{HIGHERLOOPS}.

\section{Scattering Amplitudes}
The duality relation can be extended to compute scattering amplitudes. Since the application of duality
affects only propagators, we can write down the analogue of Eq.(\ref{DualTheorem}) for amplitudes:
\begin{equation}
\label{Ampl}
{\cal A}^{(\rm{1-loop})} =-{\tilde {\cal A}}^{(\rm{1-loop})}.
\end{equation}
The expression ${\tilde {\cal A}}^{(\rm{1-loop})}$ is obtained in the following manner: start from any 
diagram in ${\cal A}^{(1-loop)}$ and consider all possible replacements of each Feynman propagator
$G_F(q_i)$ with the cut propagator. The uncut propagators are replaced by the corresponding dual propagators.
Eq.(\ref{Ampl}) establishes a correspondence between one-loop Feynman diagrams and the phase space integral of 
tree-level diagrams \cite{Catani:2008xa}. The duality relation for amplitudes is valid in any unitary and local
field theory. In spontaneously broken gauge theories, it holds 
in the 't Hooft-Feynman gauge and in the unitary gauge. 
In unbroken gauge theories, the duality 
relation is valid in the 't Hooft-Feynman gauge, and in physical 
gauges where the gauge vector $n^\nu$ is orthogonal to the 
dual vector $\eta^\mu$, i.e., $n\cdot \eta=0$. This 
excludes gauges where $n^\nu$ is time-like. This choice
of gauges at one-loop avoids the appearance of extra unphysical 
gauge poles, which in some gauges, moreover, become poles 
of higher order. At higher loop orders, the gauge poles 
will be absent with the same choice of gauges, and
the duality relation can be extended straightforwardly 
to the amplitude level too. 

For higher orders in pertubation theory, Eq.(\ref{Ampl}), generalizes to:
\begin{equation}
\label{AmplN}
{\cal A}^{(\rm{N-loop})} =(-1)^N {\tilde {\cal A}}^{(\rm{N-loop})},
\end{equation}
where ${\tilde {\cal A}}^{(\rm{N-loop})}$ is the dual counterpart of the loop quantity
${\cal A}^{(\rm{N-loop})}$, which is obtained from the Feynman graphs in ${\cal A}^{(\rm{N-loop})}$
by sequentially applying on each loop the duality theorem. In \cite{Catani:2008xa}, it was discussed
how to evaluate the on-shell scattering amplitude, of $N$ external particles, from the tree-level
forward scattering amplitude of $N+2$ particles, with two additional external legs of momenta $q$
and $-q$. The naive generalization to $N$-loops, requires the evaluation of tree-level scattering 
amplitudes with an even number of external legs, i.e. $q_i$ and $-q_i$, where $i$ runs from one to 
the number of loops (see also \cite{CaronHuot:2010zt}). Care has to be taken though, when dealing with tadpoles and self energy insertions
due to the appearence of kinematical singularities. These issues where already discussed in \cite{Catani:2008xa} 
and are left for future investigation.

\section{Conclusions}

We have rederived the tree--loop duality theorem at one--loop order, which was
introduced in Ref.~\cite{Catani:2008xa}, in a way which is more suitable for
extending it to higher loop orders. By applying iteratively the duality
theorem, we have given explicit representations of the two-- and three--loop
scalar integrals. Dual representations of the loop integrals
with complex dual prescription depending only on the external momenta can be
obtained at the cost of introducing extra cuts, which break the loop integrals
into disconnected diagrams. The maximal number of cuts agrees with the number of loop lines, and
the cuts are such that there does not appear more than a single cut for each
internal loop line.

\section*{Acknowledgments} 

This work was supported by the Ministerio de Ciencia e Innovaci\'on under
Grant No. FPA2007-60323, by CPAN (Grant No. CSD2007-00042), by the Generalitat
Valenciana under Grant No. PROMETEO/2008/069, by the European Commission under
contracts FLAVIAnet (MRTN-CT-2006-035482) and HEPTOOLS (MRTN-CT-2006-035505),
and by INFN-MICINN agreement under Grants No. ACI2009-1061 and
FPA2008-03685-E.

\end{document}